\documentclass[aps,pra,twocolumn,superscriptaddress,floatfix]{revtex4-2}
\usepackage{amsmath,amssymb,amsthm,graphicx,bm}
\usepackage{physics}
\usepackage[colorlinks=true,linkcolor=blue,citecolor=blue,urlcolor=blue]{hyperref}
\newtheorem{theorem}{Theorem}
\newtheorem{proposition}{Proposition}
\newtheorem{corollary}{Corollary}
\newtheorem{lemma}{Lemma}

\newcommand{\F}{\mathcal{F}}
\newcommand{\avg}[1]{\langle #1\rangle}

\usepackage{tikz}
\begin{document}
\title{Clifford and Haar scramblers yield equal mean fidelity but unequal fluctuations in black hole-inspired teleportation}
\author{Sudhanva Joshi}
\email{sudhanvajoshi.rs.phy24@itbhu.ac.in}
\affiliation{Department of Physics, Indian Institute of Technology (Banaras Hindu University), Varanasi 221005, India}
\author{Sunil Kumar Mishra}
\email{sunilkm.app@iitbhu.ac.in}
\affiliation{Department of Physics, Indian Institute of Technology (Banaras Hindu University), Varanasi 221005, India}

\begin{abstract}
Quantum information transfer between entangled black holes has inspired many-body teleportation protocols. We study such a protocol without assuming a gravitational dual and ask whether its fidelity requires nonstabilizerness, or magic, in the scrambling dynamics. Because the mean fidelity depends only on the third moments of the scrambler ensemble, zero-magic Clifford scramblers teleport on average as well as Haar-random unitaries. Solving the protocol exactly at infinite temperature, we find that typical Clifford scramblers approach perfect teleportation while the magic of the complete circuit vanishes as the inverse system size. The two ensembles nevertheless differ in their fluctuations: the exact Clifford fidelity variance decays only algebraically with system size, whereas the Haar variance is exponentially small. This separation characterizes the ensembles rather than magic itself, since magic added away from the message qubit can leave every fidelity statistic unchanged. With a decoder uncorrelated with the scrambler, every unitary 2-design leaves the mean fidelity at the no-transfer value, whatever its magic.
\end{abstract}

\maketitle

\section{Introduction}
\label{sec:intro}
Quantum magic (nonstabilizerness) is the resource that promotes the classically simulable stabilizer formalism to universal quantum computation~\cite{gottesman1998heisenberg,aaronson2004improved,bravyi2005universal}. The stabilizer R\'enyi entropy (SRE)~\cite{leone2022stabilizer,haug2023quantifying} provides a minimization-free measure of nonstabilizerness that is computable in many-body systems. It has since been studied in matrix-product states, fermionic Gaussian states, nonequilibrium dynamics~\cite{lami2023nonstabilizerness,collura2024quantum,paviglianiti2025estimating} and in the Sachdev-Ye-Kitaev (SYK) model~\cite{bera2025non,jasser2025stabilizer,russomanno2025nonstabilizerness,zhang2026stabilizer,sun2026connecting}. In holography, non-local magic acquires a precise geometric meaning: it vanishes if and only if gravitational back-reaction is absent, and it approximately equals the response of a minimal-surface area to the tension of a cosmic brane~\cite{cao2024non,cao2025gravitational,white2021conformal}. Back reaction of a different origin, the negative energy injected by a boundary coupling, is what makes the wormhole between the two black holes of a thermofield double traversable~\cite{gao2017traversable,maldacena2017diving} and wormhole-inspired teleportation is the operational counterpart of that traversal~\cite{gao2021traversable,brown2023quantum,nezami2023quantum,schuster2022many,jafferis2022traversable,liu2024fidelity}. It is therefore natural to ask whether the fidelity of the protocol itself draws on magic.

Schuster \textit{et al.}~\cite{schuster2022many} noted that the teleportation fidelity is built from three copies of the scrambling unitary and three of its adjoints. Because the multiqubit Clifford group forms a unitary 3-design~\cite{webb2015clifford,zhu2017multiqubit,kueng2015qubit}, random Clifford circuits reproduce the Haar-averaged fidelity, and because a Clifford scrambler maps each message operator to a single Pauli string, the fidelity of each Clifford instance follows from their peaked-size expression; they exploited both facts for efficient simulation. We treat the same observation as a statement about the resources. Below, Sec.~\ref{sec:blind} establishes that every ensemble that is a 3-design yields the identical mean fidelity, independent of the amount of magic it carries; doped Clifford circuits allow the magic to be tuned from zero to its Haar value while the mean fidelity remains fixed. Then in Sec.~\ref{sec:exact}, the infinite-temperature protocol admits an exact solution at every system size: we re-derive the fidelity of each individual Clifford scrambler from that expression and obtain a closed-form expression for the ensemble average, its thermodynamic limit, and the leading finite-size correction. In that limit, typical stabilizer scramblers teleport with unit fidelity while the magic of the full protocol vanishes; at the Clifford value of the coupling, which reduces to a register exchange, every Clifford scrambler delivers a Bell pair with no non-Clifford gates at all. Section~\ref{sec:fluct} computes the full fidelity distribution over Clifford scramblers: the Clifford and Haar ensembles, identical in the mean, differ parametrically in their fluctuations, although no fidelity statistic is a universal witness of magic. Section~\ref{sec:insufficient} shows that with an uncorrelated decoder every unitary 2-design, whatever its magic, leaves the mean fidelity exactly at the no-transfer value, and that fidelity itself supplies an upper bound on magic.

\section{Protocol}
\label{sec:protocol}
Registers ref and $M$ hold the message Bell pair, and $L$ and $R$ are $n$-qubit systems with $\mathcal{H}_L \cong \mathcal{H}_R \cong(\mathbb C^2)^{\otimes n}$ and $d=2^n$. The initial state is:
\begin{equation}
\ket{\Psi}_0 = \ket{\Phi^+}_{\rm ref,M} \otimes \ket{\Psi}_{LR},
\label{eq:init}
\end{equation}
where $\ket{\Psi}_{LR}$ may be chosen freely. Theorem~\ref{thm:main} holds for every such choice held fixed while the scrambler is sampled, including thermofield doubles (TFDs) at arbitrary inverse temperature $\beta$, but not for a scrambler $e^{-iHt}$ whose TFD is built from the same randomly drawn $H$; the exact analysis of Sec.~\ref{sec:exact} specializes to the maximally entangled state $\ket{\Phi^+}_{LR}= d^{-1/2} \sum_i \ket{i}_L\ket{i}_R$. A scrambler $U$ acting on $L$ produces the state:
\begin{equation}
\label{eq:protocol}
|\Psi_U\rangle=D_U\,C\,U\,\mathcal I\,U^\dagger\,|\Psi_0\rangle .
\end{equation}
The insertion operator $\mathcal{I}$ places the message into the first qubit $L_1$ of $L$, $C$ couples $L$ to $R$, and $D_U$ is a decoder on $R$; Fig.~\ref{fig:schematic}(a) shows the resulting circuit. $\mathcal{I}$ is either SWAP between $M$ and $L_1$, or the postselected projective insertion $\mathcal I=\tfrac12\bigl(\openone+\sum_{\sigma}\sigma_{L_1}\otimes\sigma_M^{\mathsf T}\bigr)$, with $\sigma \in \{X,Y,Z\}$, which returns the state conditioned on a successful Bell outcome, obtained with probability $1/4$. The left-right coupling is the operator-size interaction of teleportation by size~\cite{brown2023quantum}
\begin{equation}
\label{eq:coupling}
C=\exp\Bigl[ig\sum_{i=1}^{n}\bigl(\openone-\Pi_i\bigr)\Bigr],\qquad \Pi_i=|\Phi^+\rangle\langle\Phi^+|_{L_iR_i},
\end{equation}
which multiplies $O_L\ket{\Phi^+}_{LR}$ by the phase $e^{ig|O|}$ for every Pauli string $O$ of weight $|O|$. The decoder $D_U$ acts on $R$ and is taken to be either a fixed unitary or one of $U,U^\dagger,U^{\mathsf T},U^*$ under the natural identification $\mathcal{H}_R \cong \mathcal{H}_L$. The particular choice $D_U=U^{\mathsf T}$ corresponds to right-side forward evolution of the Hamiltonian protocol, since a scrambler $U=e^{-itH_L}$ is accompanied by the transposed Hamiltonian $H_R=H_L^{\mathsf T}$ on the right side~\cite{liu2024fidelity}.
The teleportation fidelity is the overlap of the reference and the first right qubit $R_1$ with the Bell state $|\Psi^-\rangle$:
\begin{equation}
\label{eq:fid}
\F(U)=\langle\Psi_U|\bigl(|\Psi^-\rangle\langle\Psi^-|\bigr)_{\mathrm{ref},R_1}|\Psi_U\rangle.
\end{equation}
In the absence of information transfer, $\F = 1/4$; the best classical (measure-and-prepare) strategy reaches $\F=1/2$, an average state fidelity of $2/3$~\cite{horodecki1999general}.

We measure the magic of a pure state of $N$ qubits by its SRE, $M_2=-\log_2\bigl(2^{-N}\sum_P\avg{P}^4\bigr)$, with $P$ running over all $4^N$ Pauli strings~\cite{leone2022stabilizer}. The magic $M_2(W)$ of a unitary $W$ is the SRE of its normalized Choi state $(W\otimes\openone)\ket{\Phi^+}$; it vanishes if and only if $W$ is Clifford, is additive under tensor products, and characterizes the circuit rather than any state it produces. While magic monotones such as the SRE furnish lower bounds on the non-Clifford cost of a circuit~\cite{beverland2020lower}, they do not supply matching upper bounds, which is a distinction that becomes essential in Sec.~\ref{sec:exact}.

\begin{figure}[t]
\centering
\definecolor{scr}{RGB}{106,81,163}\definecolor{cpl}{RGB}{217,95,2}
\definecolor{cliff}{RGB}{203,24,29}\definecolor{haar}{RGB}{8,48,107}
\begin{tikzpicture}[font=\small,
  w/.style={line width=0.6pt},
  g/.style={draw=scr, fill=scr!12, line width=0.6pt, minimum width=0.66cm, minimum height=0.48cm, inner sep=1pt},
  c/.style={draw=cpl, fill=cpl!15, line width=0.6pt, minimum width=0.66cm, minimum height=1.18cm, inner sep=1pt}]
\def\yA{0}\def\yM{-0.62}\def\yL{-1.32}\def\yR{-2.02}
\node[font=\normalsize] at (-0.62,0.42) {(a)};
\foreach \y/\t in {\yA/ref,\yM/$M$,\yL/$L$,\yR/$R$} \node[anchor=east] at (-0.08,\y) {\t};
\draw[w] (0.34,\yA) .. controls (0.02,\yA) and (0.02,\yM) .. (0.34,\yM);
\draw[w] (0.34,\yL) .. controls (0.02,\yL) and (0.02,\yR) .. (0.34,\yR);
\node[font=\footnotesize] at (0.74,{(\yA+\yM)/2}) {$|\Phi^+\rangle$};
\node[font=\footnotesize] at (0.78,{(\yL+\yR)/2}) {$|\Psi_{LR}\rangle$};
\draw[w] (0.34,\yA) -- (6.30,\yA);  \draw[w] (0.34,\yM) -- (5.78,\yM);
\draw[w] (0.34,\yL) -- (5.78,\yL);  \draw[w] (0.34,\yR) -- (6.30,\yR);
\foreach \y/\d in {\yL/0.19,\yR/-0.19} {\draw[w] (1.26,\y-0.09) -- (1.38,\y+0.09); \node[font=\scriptsize] at (1.47,\y+\d) {$n$};}
\node[g] at (1.95,\yL) {$U^{\dagger}$};
\draw[w] (2.72,\yM) -- (2.72,\yL);
\foreach \y in {\yM,\yL} {\draw[w] (2.64,\y-0.08) -- (2.80,\y+0.08); \draw[w] (2.64,\y+0.08) -- (2.80,\y-0.08);}
\node[font=\footnotesize,anchor=west] at (2.76,{(\yM+\yL)/2}) {$\mathcal I$};
\node[font=\scriptsize] at (2.72,\yL-0.24) {$L_1$};
\node[g] at (3.50,\yL) {$U$};
\node[c] at (4.40,{(\yL+\yR)/2}) {$C$};
\node[g] at (5.30,\yR) {$U^{\mathsf T}$};
\node[font=\scriptsize] at (6.08,\yR+0.19) {$R_1$};
\draw[w] (6.30,\yA+0.13) -- (6.44,\yA+0.13) -- (6.44,\yR-0.13) -- (6.30,\yR-0.13);
\node[anchor=west] at (6.48,{(\yA+\yR)/2+0.14}) {$\mathcal F$};
\node[anchor=west,font=\footnotesize] at (6.48,{(\yA+\yR)/2-0.26}) {$|\Psi^-\rangle$};
\def\yb{-4.70}\def\hb{1.05}
\node[font=\normalsize] at (-0.62,-2.88) {(b)};
\node[font=\footnotesize] at (3.40,-2.88) {size distribution of $U\sigma_{L_1}U^{\dagger}$};
\foreach \x/\col/\tit/\mag in {0.25/cliff/{Clifford: one Pauli string}/{$M_2=0$},
                                 3.95/haar/{Haar: many Pauli strings}/{$M_2\approx M_2^{\rm Haar}$}} {
  \node[font=\footnotesize,text=\col] at (\x+1.45,-3.24) {\tit};
  \draw[w] (\x,\yb) -- (\x+2.90,\yb);
  \node[font=\scriptsize,below] at (\x+0.12,\yb) {$0$};
  \node[font=\scriptsize,below] at (\x+2.78,\yb) {$n$};
  \node[font=\scriptsize,below] at (\x+1.45,\yb) {size $s$};
  \node[font=\footnotesize,text=\col,anchor=north west] at (\x+0.02,-3.48) {\mag};
}
\fill[cliff] (0.25+0.12+6*0.333-0.11,\yb) rectangle (0.25+0.12+6*0.333+0.11,\yb+\hb*0.92);
\foreach \s/\h in {1/0.001,2/0.012,3/0.074,4/0.278,5/0.667,6/1.0,7/0.857,8/0.321}
  \fill[haar] (3.95+0.12+\s*0.333-0.11,\yb) rectangle (3.95+0.12+\s*0.333+0.11,\yb+\hb*0.92*\h);
\node[font=\small] at (3.40,-5.45) {$\langle\mathcal F\rangle_{\rm Clifford}=\langle\mathcal F\rangle_{\rm Haar}$};
\end{tikzpicture}
\caption{(a)~Protocol of Eq.~\eqref{eq:protocol}. Registers $\mathrm{ref}$ and $M$ hold a Bell pair $|\Phi^+\rangle$; the $n$-qubit systems $L$ and $R$ begin in the state $|\Psi_{LR}\rangle$. The scrambler $U$ is applied to $L$ both before and after the insertion $\mathcal{I}$ (shown here as the SWAP of $M$ with the first left qubit $L_1$). The coupling $C$ of Eq.~\eqref{eq:coupling} multiplies $O_L|\Phi^+\rangle_{LR}$ by the phase $e^{ig|O|}$ for every Pauli string $O$. The decoder $U^{\mathsf{T}}$, the right-side image of the scrambler, acts on $R$. Teleportation fidelity $\mathcal{F}$ (Eq.~\eqref{eq:fid}) is the overlap of the reference and the first right qubit $R_1$ with the singlet $|\Psi^-\rangle$. The three operations constructed from the same unitary $U$ are drawn in a common color. (b)~Size distribution of the scrambled message operator. A Clifford scrambler maps $\sigma_{L_1}$ onto a single Pauli string of definite weight, and for $|\Psi_{LR}\rangle=|\Phi^+\rangle_{LR}$ the scrambled state has $M_2=0$. A Haar-random scrambler distributes the same operator over many Pauli strings (the bars show the Haar-averaged size distribution for $n=8$, normalized to its maximum) and produces a scrambled state whose SRE is close to its mean $M_2^{\rm Haar}$ over Haar-random scramblers. Despite the disparity in magic, the mean fidelities of the two ensembles coincide (Theorem~\ref{thm:main}).}
\label{fig:schematic}
\end{figure}

\section{The mean fidelity is blind to magic}
\label{sec:blind}

\begin{lemma}
\label{lem:norm}
$\|\mathcal I\,U^\dagger|\Psi_0\rangle\|=1$ for every $U$.
\end{lemma}
\begin{proof}
The claim is immediate when $\mathcal{I}$ is a SWAP. For the projective insertion, one has $\mathcal{I}=2\ket{\Phi^+}\bra{\Phi^+}$, and therefore $\mathcal{I}^\dagger \mathcal{I}=2\mathcal{I}$. Prior to insertion, the register $M$ is maximally mixed and uncorrelated with $L$, so $\avg{\mathcal I}=\tfrac12\bigl[1+\sum_\sigma\avg{\sigma_{L_1}}\avg{\sigma_M^{\mathsf T}}\bigr]=\tfrac12$ and $\avg{\mathcal I^\dagger\mathcal I}=1$.
\end{proof}

\begin{theorem}[Design invariance, observed in Ref.~\cite{schuster2022many}]
\label{thm:main}
With the resource state, insertion, coupling, and decoder rule fixed independently of the sampled $U$, the fidelity $\F(U)$ is a homogeneous polynomial of bidegree $(3,3)$ in the matrix elements of $(U,U^*)$. When the decoder $D_U$ is independent of $U$, the bidegree drops to $(2,2)$. Consequently,
\begin{equation}
\label{eq:main}
\mathop{\mathbb E}_{U\sim\mathcal E}\F(U)=\mathop{\mathbb E}_{U\sim\mathrm{Haar}}\F(U)
\end{equation}
for every ensemble $\mathcal{E}$ that forms a unitary 3-design (and for every unitary 2-design when $D_U$ is fixed).
\end{theorem}

\begin{proof}
In the protocol of Eq.~\eqref{eq:protocol}, the scrambler $U$ and its adjoint each appear once, contributing matrix elements of $U$ and $U^*$. The decoder $D_U$ can introduce at most one further factor of $U$ or $U^*$. The resulting state vector $\ket{\Psi_U}$ is therefore homogeneous of bidegree at most $(2,1)$ or $(1,2)$; its dual carries the transposed bidegree, and the fidelity $\F$ is homogeneous of bidegree at most $(3,3)$. Lemma~\ref{lem:norm} guarantees that no $U$-dependent normalisation intervenes. A unitary $t$-design reproduces the Haar average of $U^{\otimes t}\otimes U^{*\otimes t}$, and therefore the average of every homogeneous polynomial of bidegree $(t,t)$.
\end{proof}

Schuster \textit{et al.} made this counting observation and used it to efficiently simulate random-circuit scrambling~\cite{schuster2022many}; we state it as a theorem to make the resource content explicit: mean fidelity is insensitive to any feature of the ensemble that lies beyond its third moments.

\begin{corollary}
\label{cor:clifford}
(1) The $n$-qubit Clifford group $\mathcal{C}_n$ is a unitary 3-design~\cite{webb2015clifford,zhu2017multiqubit,kueng2015qubit} consisting entirely of zero-magic unitaries. Hence, $\mathbb E_{\mathcal C_n}\F=\mathbb E_{\rm Haar}\F$. When the initial state $\ket{\Psi_{LR}}$ is a stabilizer state (in particular at $\beta=0$), the intermediate state $U\mathcal IU^\dagger|\Psi_0\rangle$ itself has vanishing $M_2$ for every $U \in \mathcal{C}_n$.

(2) For any $t \geq 0$, the doped Clifford ensemble
\begin{equation}
\label{eq:doped}
U=C_t\,T\,C_{t-1}\,T\cdots T\,C_0 ,
\end{equation}
With independent uniform Clifford factors $C_k$ and $T$ gates on sites chosen independently of them, it remains an exact unitary 3-design, so its mean fidelity coincides with the Haar value irrespective of the number of non-Clifford gates.
\end{corollary}
\begin{proof}
Part (1) follows from Theorem~\ref{thm:main} together with the Clifford invariance of $M_2$~\cite{leone2022stabilizer}. For part (2), write $U=C_tV$ with $V$ independent of the outermost Clifford. Because $C_t$ is drawn from a 3-design, one has:
\begin{equation}
\mathbb E[(C_tV)^{\otimes3}\otimes(C_tV)^{*\otimes3}]=\mathbb E_{W\sim{\rm Haar}}[(WV)^{\otimes3}\otimes(WV)^{*\otimes3}]
\end{equation}
for each fixed $V$. Invariance of Haar measure then implies that $W V$ is itself Haar-distributed.
\end{proof}
Figure~\ref{fig:schematic}(b) illustrates part (1). A Clifford scrambler sends the message operator to a single Pauli string; a Haar-random scrambler spreads it over many; and the two yield the same mean fidelity.

The SRE of a state $U \ket{\phi}$ is controlled by the fourth-moment quantity $d^{-1}\sum_P\langle\phi|U^\dagger PU|\phi\rangle^4$; for the intermediate state $U\mathcal IU^\dagger|\Psi_0\rangle$, which contains $U$ twice, moments up to eighth order enter. The Clifford group is not a 4-design~\cite{webb2015clifford,zhu2016clifford}; its fourth moments depart from the Haar ones through the additional elements of the Clifford commutant~\cite{gross2021schur,bittel2026complete}, which is the structure the SRE detects~\cite{leone2022stabilizer}. The mean teleportation fidelity, a third-moment quantity, is blind to that departure; its variance, which involves sixth moments, is not, as Sec.~\ref{sec:fluct} shows for the Clifford and Haar ensembles.

\section{Exact solution at infinite temperature}
\label{sec:exact}
Henceforth, $|\Psi_{LR}\rangle=|\Phi^+\rangle_{LR}$, the insertion is SWAP, and $D_U=U^{\mathsf T}$. (Theorem~\ref{thm:main} continues to apply to the Majorana double-trace coupling of the SYK protocol~\cite{gao2021traversable}; an exact treatment of that case would require tracking Majorana rather than Pauli weight and is left aside.) For a Clifford scrambler, let $w_\sigma$ denote the Pauli weight of the conjugated operator $U\sigma_{L_1}U^\dagger$, with $\sigma\in\{X,Y,Z\}$.

\begin{proposition}
\label{prop:exact}
(1) The fidelity of an arbitrary Clifford scrambler is
\begin{equation}
\label{eq:instance}
\F=\frac1{16}\Bigl|1-e^{igw_X}-e^{igw_Y}-e^{igw_Z}\Bigr|^2 .
\end{equation}
(2) For any ensemble that forms a unitary 3-design (in particular for Haar-random unitaries), the average fidelity admits the closed form
\begin{equation}
\label{eq:closed}
\avg{\F}=\frac14+\frac38\bigl(B_n-\mathrm{Re}\,A_n\bigr),
\end{equation}
where
\begin{align}
A_n&=\frac{(1+3e^{ig})^n-1}{4^n-1},\label{eq:AB}\\
B_n&=\frac{(10+6\cos g)^n-(6\cos g-2)^n}{16^n-4^n}.\nonumber
\end{align}

\end{proposition}
Part~(1) is the single-qubit case of the peaked-size fidelity of Schuster \textit{et al.} [Eq.~(44) of Ref.~\cite{schuster2022many}], which they evaluated with phases set by operator sizes to simulate Clifford circuits; Appendix~\ref{app:derivation} gives a self-contained derivation. Eq.~\eqref{eq:instance} shows that the fidelity of a Clifford instance is completely determined by three non-negative integers; the overall signs of the conjugated Paulis cancel. Part~(2) is obtained by averaging Eq.~\eqref{eq:instance} over the uniform Clifford group under which the images of $X$ and $Z$ form a uniformly random anti-commuting pair of Pauli strings, post which Theorem~\ref{thm:main} is invoked. The resulting expression was verified by exhaustive enumeration of all $(16^n-4^n)/2$ anticommuting pairs for every $2 \leq n \leq 5$, agreeing to machine precision.

\begin{corollary}
\label{cor:limit}
We set $g=\gamma/n$. In the thermodynamic limit
\begin{equation}
\label{eq:limit}
\avg{\F}\to\frac14+\frac38\Bigl[1-\cos\bigl(\tfrac34\gamma\bigr)\Bigr],
\end{equation}
which attains unity at $\gamma = 4\pi/3$. At that coupling strength, the leading finite-size correction is
\begin{equation}
\label{eq:finite}
1-\avg{\F}=\frac{3\pi^2}{16\,n}+O(n^{-2}).
\end{equation}
When the scrambler is Clifford, the Choi-state magic of the full protocol unitary coincides with that of the coupling alone,
\begin{equation}
\label{eq:magicexact}
M_2(C)=-n\log_2\Bigl[\frac{89+36\cos 2g+3\cos 4g}{128}\Bigr],
\end{equation}
which at $\gamma=4\pi/3$ becomes
\begin{equation}
\label{eq:magiccoupling}
M_2(C)=\frac{4\pi^2}{3\,n\ln2}+O(n^{-3}).
\end{equation}

\end{corollary}

The limiting expression Eq.~\eqref{eq:limit} is the peaked-size fidelity of~\cite{schuster2022many} with every message operator at the typical Haar size $3n/4$, and Eq.~\eqref{eq:finite} makes exact the $1/n$ scaling of finite-size errors estimated there. Because $\F \leq1$, an average that approaches unity forces typical instances to approach unity as well, so typical stabilizer scramblers teleport perfectly in the large-$n$ limit. The identification $M_2(\text{protocol}) = M_2(C)$ follows because, with $D_U=U^{\mathsf T}$, the complete evolution is a product of the coupling $C$ with Clifford unitaries on either side, and the Choi-state SRE is invariant under such multiplications. Combining Eq.~\eqref{eq:finite} and Eq.~\eqref{eq:magiccoupling} yields the proportional relation $M_2(C)=\frac{64}{9\ln2}(1-\avg{\F})+O(n^{-2})$ along the asymptotically optimal coupling $g=4\pi/(3n)$: perfect teleportation is approached while the magic of the protocol vanishes linearly with the residual infidelity.

Eq.~\eqref{eq:magiccoupling} quantifies magic by the SRE, and a small-angle rotation carries a small SRE even though an accurate Clifford+$T$ synthesis of the same unitary may demand many non-Clifford gates. The following corollary is free of that subtlety.

\begin{corollary}
\label{cor:pi}
At $g=\pi$, the coupling itself is Clifford. Consequently, when the scrambler is also Clifford, the entire circuit is a stabilizer circuit containing no non-Clifford gate and is efficiently classically simulable~\cite{gottesman1998heisenberg,aaronson2004improved}. Every instance yields a fidelity $\in \{0,1\}$, and the fraction of Clifford instances with $\F=1$ is
\begin{equation}
\label{eq:pi}
\Pr(\F=1)=\frac14-\frac34\,\frac{(-2)^n-1}{4^n-1}\;\longrightarrow\;\frac14 .
\end{equation}
\end{corollary}
The coupling factor $2\Pi_i-\openone$ equals $-(Y\otimes Y)\,\mathrm{SWAP}_{L_iR_i}$, so at $g=\pi$ the coupling exchanges $L$ and $R$ up to local Paulis. Every instance then leaves ref and $R_1$ in a pure Bell state; the instances with $\F=0$ deliver one of the other three Bell states, which a Pauli correction determined by $U$ removes. This is a strong-coupling register exchange, not the weak-coupling scrambling mechanism of Corollary~\ref{cor:limit}. The parity argument is given in Appendix~\ref{app:derivation}.

\section{Fluctuations distinguish Clifford from Haar scramblers}
\label{sec:fluct}
Theorem~\ref{thm:main} determines only the mean fidelity. Because the fidelity of any Clifford instance depends solely on the triple $(w_X,w_Y,w_Z)$, every moment of its distribution over the Clifford group is fixed by the joint characteristic function of the three weights.

\begin{proposition}
\label{prop:fluct}
(1) Over the uniform Clifford group,
\begin{equation}
\label{eq:charfun}
\mathbb E\,e^{i(\alpha w_X+\beta w_Y+\delta w_Z)}=\frac{t_+^{\,n}-t_-^{\,n}}{16^n-4^n},
\end{equation}
where
\begin{equation}
t_\pm=1+3e^{i(\beta+\delta)}+3e^{i(\alpha+\beta)}+3e^{i(\alpha+\delta)}\pm6e^{i(\alpha+\beta+\delta)}.
\end{equation}
Expanding $16^{-k}|1-e^{igw_X}-e^{igw_Y}-e^{igw_Z}|^{2k}$ in these phases yields $\mathbb E\,\F^k$ exactly for every $k$ and $n$; the case $k=1$ recovers Eq.~\eqref{eq:closed}. (2) With $g=\gamma/n$,
\begin{equation}
\label{eq:varcliff}
\mathrm{Var}_{\mathcal C_n}\F=\frac{9\gamma^2\sin^2(3\gamma/4)}{1024\,n}+O(n^{-2}),
\end{equation}
and at $\gamma=4\pi/3$, where the leading term vanishes, $\mathrm{Var}_{\mathcal C_n}\F=11\pi^4/(384\,n^2)+O(n^{-3})$.
(3) For $n\ge2$ and Haar-random scramblers, $\mathrm{Var}_{\rm Haar}\F\le1728/(2^n-2)$.
\end{proposition}

\begin{proof}
Part~(1) extends the counting of Appendix~\ref{app:derivation} to three phases, with $y$ the indicator of a non-identity entry of $P_XP_Z$ on a given site. For part~(2), a second-order expansion of Eq.~\eqref{eq:charfun} shows that the rescaled weights $(w_\sigma-3n/4)/\sqrt n$ converge to independent Gaussian variables of mean zero and variance $3/16$; expanding $\F$ about $gw_\sigma=3\gamma/4$ then produces the variance (Appendix~\ref{app:derivation}). For part~(3), the SWAP insertion together with $D_U=U^{\mathsf T}$ gives $|\F(U)-\F(V)|\le2\|(W_U-W_V)|\Psi_0\rangle\|\le6\|U-V\|_{\rm HS}$, where $W_U=D_U\,C\,U\,\mathcal I\,U^\dagger$. Concentration of measure on the unitary group~\cite{meckes2019random} then implies $\Pr(|\F-\mathbb E\F|\ge\epsilon)\le2e^{-(2^n-2)\epsilon^2/864}$; integrating over $\epsilon^2$ recovers the stated bound.
\end{proof}

The two ensembles realize the same mean infidelity by opposite mechanisms. A Clifford scrambler assigns each message operator a definite size, so the entire mean infidelity arises from the spread of sizes across instances: at $\gamma=4\pi/3$, independent Gaussian weights give $\mathbb E(1-\F)=\tfrac9{16}g^2\,\mathrm{Var}\,w=27\gamma^2/256n=3\pi^2/16n$, which is Eq.~\eqref{eq:finite}. A Haar scrambler, by contrast, endows every instance with the full size distribution; the same mean infidelity is then generated by the width of that distribution inside each instance, while the variation between instances is suppressed to exponentially small size by concentration of measure. Consequently, the spread of the fidelity for Clifford scramblers is comparable to the infidelity itself, with a ratio $(11\pi^4/384)^{1/2}/(3\pi^2/16)\approx0.90$ at $\gamma=4\pi/3$. Fig.~\ref{fig:main}(b) makes the separation visible: a fit over $n=3$ to $8$ gives a Haar variance decaying as $2^{-(2.20\pm0.02)n}$, faster than the bound of part~(3), which is loose and becomes nontrivial only for $n\ge13$; at $n=8$ the Clifford variance, $0.0208$, exceeds the Haar variance by a factor of $4\times10^4$. Doped circuits interpolate between the two extremes, as the narrowing band of Fig.~\ref{fig:main}(c) illustrates. Contrasts between Clifford and Haar dynamics in quantities beyond the third moment are already known for out-of-time-order correlators~\cite{yoshida2019disentangling,leone2021quantum}; here, the difference is invisible in the first moment of an operational figure of merit and appears in the second.

The separation is a property of these two ensembles, not a calibration of fidelity statistics against magic. A unitary $V$ acting only on $L_2,\dots,L_n$ commutes with the insertion, so $(UV)\mathcal I(UV)^\dagger=U\mathcal IU^\dagger$, and with the matched decoder $|\Psi_{UV}\rangle=V^{\mathsf T}_R|\Psi_U\rangle$; since $V^{\mathsf T}_R$ acts away from $R_1$, $\F(UV)=\F(U)$ for every $U$. Right-multiplying a uniformly random Clifford scrambler by $V=\openone\otimes T^{\otimes(n-1)}$ therefore leaves every fidelity statistic unchanged while raising the Choi-state magic of the scrambler from zero to $(n-1)\log_2(4/3)$.

\section{Magic is not sufficient, and fidelity bounds magic}
\label{sec:insufficient}
What fidelity tracks is the correlation between the scrambler and the decoder. When the decoder is matched, $D_U=U^{\mathsf T}$, every unitary 3-design, whether of vanishing magic or of Haar-typical magic, attains the same average fidelity provided in Eq.~\eqref{eq:closed}. The moment the decoder is chosen independently of the scrambler, that average collapses to the no-transfer value.

\begin{proposition}
\label{prop:fixed}
For either insertion, any fixed resource state and coupling, and a decoder $D$ independent of $U$, every unitary 2-design gives $\mathbb E_U\F=1/4$, whatever its magic.
\end{proposition}
\begin{proof}
Let $P$ be a Pauli operator on $L_1$. Both insertions satisfy $P_{L_1}\mathcal IP_{L_1}=P_M\mathcal IP_M$, and $P_M$ commutes with $U$, $C$, and $D$, so the circuit for the scrambler $UP$ is $P_MW_UP_M$, with $W_U=D\,C\,U\mathcal IU^\dagger$. Since $P_M|\Phi^+\rangle_{\mathrm{ref},M}=P^{\mathsf T}_{\mathrm{ref}}|\Phi^+\rangle_{\mathrm{ref},M}$, this gives $|\Psi_{UP}\rangle=P_MP^{\mathsf T}_{\mathrm{ref}}|\Psi_U\rangle$: the reduced state of ref and $R_1$ for $UP$ is that for $U$, conjugated by a Pauli on ref. The Haar measure is invariant under $U\to UP$, so averaging over the four Paulis twirls the reference, $\mathbb E\,\rho_{\mathrm{ref},R_1}=\tfrac12\openone\otimes\bar\rho_{R_1}$, whose overlap with every Bell state is $1/4$. Theorem~\ref{thm:main} transfers this value to every unitary 2-design.
\end{proof}

At the asymptotically optimal coupling $\gamma=4\pi/3$ and with a fixed Haar-random decoder, sampling confirms this:
\begin{align*}
n=3:\quad
&\avg{\F}=0.2486\pm0.0008\ \text{(Clifford)},\\ 
&\avg{\F}=0.2489\pm0.0008\ \text{(Haar)},\\[4pt]
n=4:\quad 
&\avg{\F}=0.2500\pm0.0004\ \text{(Clifford)},\\
&\avg{\F}=0.2502\pm0.0005\ \text{(Haar)}.
\end{align*}
These values contrast with the matched-decoder values $0.597$ and $0.663$, respectively. Thus, even a scrambler that carries the full Haar-typical magic has a mean fidelity at the no-transfer value once its decoder is uncorrelated with the scrambler. High fidelity also imposes an upper bound on magic.

\begin{proposition}
\label{prop:bound}
Let $\F=\tfrac14\bigl(1+\sum_{\sigma}\lambda_\sigma\avg{\sigma\otimes\sigma}\bigr)$ with $\lambda_\sigma=\pm1$ be the fidelity with an arbitrary Bell state, evaluated on a pure state of $N$ qubits. Then,
\begin{equation}
\label{eq:bound}
M_2\;\le\;N-\log_2\Bigl[1+\tfrac1{27}\,(4\F-1)^4\Bigr].
\end{equation}
\end{proposition}
\begin{proof}
Writing $a,b,c$ for three Pauli expectation values and setting $s=4\F-1=\lambda_Xa+\lambda_Yb+\lambda_Zc$, Cauchy-Schwarz gives $s^2\le3(a^2+b^2+c^2)$. The power-mean inequality further gives $a^4+b^4+c^4\ge\tfrac13(a^2+b^2+c^2)^2\ge s^4/27$. Including the identity contribution, one obtains $\sum_P\avg{P}^4\ge1+s^4/27$. Substituting into the definition of $M_2$ produces the claimed bound.
\end{proof}
At perfect fidelity $\F=1$, the inequality reduces to $M_2\le N-2$, consistent with the exact structure at that point: purity of the Bell marginal forces $|\Psi\rangle=|\mathrm{Bell}\rangle_{\mathrm{ref},R_1}\otimes|\chi\rangle$, so $M_2(\Psi)=M_2(\chi)$ and the teleported pair carries no SRE. Eq.~\eqref{eq:bound} quantifies how this constraint relaxes at imperfect fidelity.

\section{Numerical confirmation}
\label{sec:numerics}
\begin{figure*}
\centering
\includegraphics[width=\textwidth]{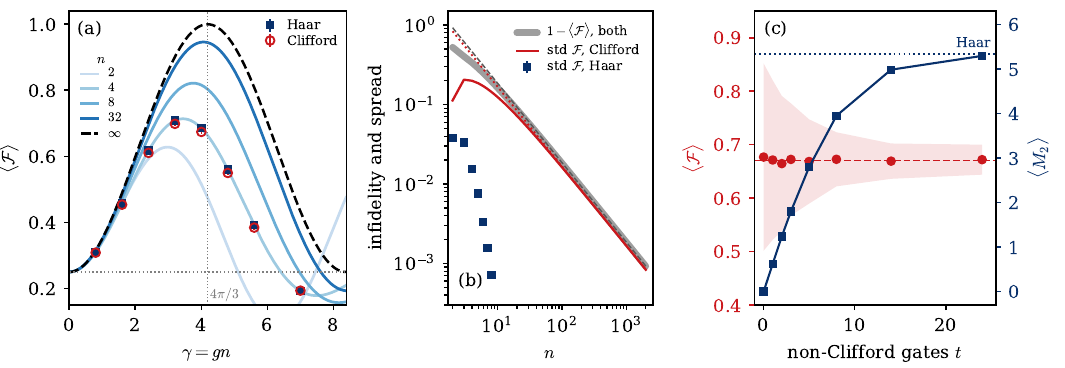}
\caption{(a) Mean fidelity versus $\gamma=gn$ obtained from the closed form Eq.~\eqref{eq:closed} for system sizes $n=2,4,8,32$ (solid curves, light to dark) together with the thermodynamic limit Eq.~\eqref{eq:limit} (dashed), which reaches unity at $\gamma=4\pi/3$ (vertical dotted line). Open circles and filled squares show Monte-Carlo averages over $1000$ uniformly random Clifford and Haar-random scramblers, respectively, at $n=4$; error bars are standard errors of the mean, and every point lies within $1.8$ standard errors of the analytic curve. The horizontal dotted line marks the no-transfer value $1/4$. (b) At $\gamma=4\pi/3$: mean infidelity $1-\avg{\F}$, identical for both ensembles (gray, exact), standard deviation of $\F$ over Clifford scramblers (red, exact, Proposition~\ref{prop:fluct}), and over Haar-random scramblers (squares, direct simulation with bootstrap errors). The dashed and dotted lines are the asymptotes $3\pi^2/16n$ and $(11\pi^4/384)^{1/2}/n$. (c) Doped Clifford ensemble Eq.~\eqref{eq:doped} at $n=3$ and $g=1.095$, the coupling that maximizes Eq.~\eqref{eq:closed} at this size, $1500$ samples per doping level $t$. Mean fidelity (circles, left axis) remains locked to the closed-form value $0.6705$ (dashed) for every $t$, while the SRE of the scrambled state $U\mathcal IU^\dagger|\Psi_0\rangle$ on all $2n+2$ qubits (squares, right axis) climbs from zero toward $5.33$ (dotted), its mean for Haar-random scramblers in the same protocol. The shaded band indicates $\pm1$ standard deviation of the fidelity distribution and narrows from $0.175$ to $0.028$.}
\label{fig:main}
\end{figure*}
Figure~\ref{fig:main} summarizes the principal numerical tests. Direct evaluation of the protocol for sixty independent Clifford scramblers at each of $n=2,3,4,5$ and nine values of coupling reproduces Eq.~\eqref{eq:instance} to ($10^{-7}$), the single-precision floor of the Clifford unitaries. Monte-Carlo averages constructed from 3000 Clifford and 3000 Haar samples at $n=3$, and from $1500$ samples of each ensemble at $n=4$ agree with the closed-form expression Eq.~\eqref{eq:closed} to within 2.2 standard errors across all thirty-six comparisons (nine couplings $\times$ two ensembles $\times$ two sizes). An independent sample at $n=4$ is displayed in panel~(a); panel~(b) shows the exact mean infidelity and Clifford spread approaching their asymptotes, and at $n=4096$ the closed form agrees with Eq.~\eqref{eq:finite} to better than $0.1\%$. The exact second moment of Proposition~\ref{prop:fluct} agrees with exhaustive enumeration to $10^{-15}$ for $2\le n\le5$.

At the special point $g=\pi$, every sampled Clifford instance returns a fidelity belonging to the set $\{0,1\}$ for system sizes $n=2$ through $6$, and the observed frequencies are statistically consistent with the exact fraction Eq.~\eqref{eq:pi}. Panel~(c) examines the doped Clifford family. For every doping level $t$, the sample mean of the fidelity remains inside 2.1 standard errors of the analytic value $0.6705$, while the SRE of the scrambled state rises from $0.000$ to $5.29$, approaching $5.33$, its mean with Haar-random scramblers.

Theorem~\ref{thm:main} is independent of inverse temperature. When the left-right state is taken to be the TFD of a fixed random Hamiltonian at $n=3$, the Clifford and Haar averages coincide within $2.0$ standard errors at each of the four temperatures $\beta = 0,1,2,4$ and at both couplings examined.

Theorem~\ref{thm:main} and Proposition~\ref{prop:fixed} apply equally to a scrambler acting on the whole left-right register, since their proofs use only the number of occurrences of $U$ and its invariance under Paulis on $L_1$. Explicit Monte-Carlo sampling in an SYK implementation, where a random Clifford circuit replaces the SYK scrambler on the full left-right register and the SYK right-side evolution serves as the fixed decoder (30 realizations of each ensemble; Appendix~\ref{app:numerics}), confirms the prediction: Clifford and Haar fidelities differ by at most $0.002$ at every extraction time, with the largest deviation being 1.6 standard errors, and both averages sit at the no transfer value of $1/4$. The absence of teleportation is therefore a consequence of the decoder-scrambler mismatch and not of the vanishing magic of the Clifford ensemble.

\section{Discussion}
\label{sec:discussion} 
Theorem~\ref{thm:main} answers the question of Sec.~\ref{sec:intro} in the negative at the level of the scrambler: the mean fidelity is the same for ensembles that carry no magic and for ensembles that carry near-maximal magic. The conclusion is consistent with the observation of Schuster \textit{et al.} that peaked-size teleportation requires no gravitational dual~\cite{schuster2022many}. Within the infinite-temperature protocol studied here, a high teleportation fidelity of the kind reported on quantum hardware~\cite{jafferis2022traversable} therefore does not, by itself, certify nonstabilizerness: typical stabilizer scramblers already reach unit fidelity with a coupling whose Choi-state magic vanishes as $1/n$ (Corollary~\ref{cor:limit}). This concerns the magic of the circuit, not the non-local magic of holographic states, which~\cite{cao2025gravitational} relates to gravitational back-reaction. The fidelity distribution does separate the two ensembles: its spread differs between stabilizer and Haar-random scramblers by a factor that grows exponentially with $n$ (Proposition~\ref{prop:fluct}). This is a property of those ensembles rather than a universal witness of magic, since the spectator construction of Sec.~\ref{sec:fluct} adds extensive magic to the scrambler without changing any fidelity statistic.

The main limitation is the gravitational regime. Theorem~\ref{thm:main} guarantees the Haar mean for unitary 3-design ensembles, or for unitary 2-design ensembles with a fixed decoder, provided the resource state is held fixed independently of the sampled scrambler. These sufficient conditions are not automatically satisfied by evolution and thermal states generated from the same random SYK Hamiltonian. At low temperature, the scrambler remains correlated with the thermal state, and the teleportation proceeds by size-winding~\cite{nezami2023quantum,brown2023quantum}: the size distribution of the thermally dressed message operator acquires size-dependent phases. At infinite temperature, a Clifford scrambler maps each message operator into a single Pauli string so the scrambled operator has no size distribution that could wind. Whether the magic of a structured, energy-conserving scrambler becomes essential in that low-temperature regime is left open by the present analysis.

\section*{Data Availability}
The data supporting the findings of this paper are openly available at \cite{joshi2025witp}.

\appendix
\begin{widetext}
\section{Derivations}
\label{app:derivation}
\paragraph{Fidelity of a Clifford scrambler.}
Expand the insertion as $\mathrm{SWAP}_{M,L_1}=\tfrac12\sum_{\sigma}\sigma_M\otimes\sigma_{L_1}$ with $\sigma\in\{\openone,X,Y,Z\}$. Conjugation by the scrambler then yields: $U\,\mathrm{SWAP}\,U^\dagger=\tfrac12\sum_\sigma\sigma_M\otimes(U\sigma_{L_1}U^\dagger)$. Acting on the reference Bell pair, one has $\sigma_M|\Phi^+\rangle_{\mathrm{ref},M}=\sigma^{\mathsf T}_{\mathrm{ref}}|\Phi^+\rangle_{\mathrm{ref},M}$. When $U$ is Clifford, then each conjugated operator $U\sigma_{L_1}U^\dagger$ is a single signed Pauli string of weight $w_\sigma$ (with convention $w_{ \openone} = 0$), and the coupling multiplies the corresponding term by the phase $e^{igw_\sigma}$. The identity $(A_R)|\Phi^+\rangle_{LR}=(A^{\mathsf T})_L|\Phi^+\rangle_{LR}$ together with the decoder $U_R^{\mathsf T}$ produces $U^{\mathsf T}_R\,(U\sigma U^\dagger)_L|\Phi^+\rangle_{LR}=(U\sigma)_L|\Phi^+\rangle_{LR}$. Collecting all factors, one obtains:
\begin{equation}
|\Psi_U\rangle=\tfrac12\textstyle\sum_\sigma e^{igw_\sigma}\,\sigma^{\mathsf T}_{\mathrm{ref}}\otimes(U\sigma_{L_1})_L\,|\Phi^+\rangle_{\mathrm{ref},M}|\Phi^+\rangle_{LR}.
\end{equation}
After the operators on $L$ are moved to $R$ and the registers $M,L$ and all of $R$ except $R_1$ are traced out, the factors of $U$ cancel by virtue of $U^{\mathsf T}U^* = \openone$. The resulting reduced state is $\rho_{\mathrm{ref},R_1}=KK^\dagger$ with
\begin{equation}
K=\tfrac14\sum_\sigma e^{igw_\sigma}\sigma\otimes\sigma
\end{equation}
(using the transposition identity $\sigma^{\mathsf T}\otimes\sigma^{\mathsf T}=\sigma\otimes\sigma$). The operator $K$ is diagonal in the Bell basis; the singlet $\ket{\Psi^-}$ is the simultaneous eigenstate of $X\otimes X$, $Y\otimes Y$, and $Z\otimes Z$ with eigenvalue $-1$. Evaluating this overlap gives Eq.~\eqref{eq:instance}.
\paragraph{Ensemble average.}
Expanding the exact expression Eq.~\eqref{eq:instance} with $\varphi_\sigma=e^{igw_\sigma}$ produces
\begin{equation}
\F=\tfrac1{16}\bigl[4-2\,\mathrm{Re}\sum_\sigma\varphi_\sigma+2\,\mathrm{Re}\sum_{\sigma<\tau}\varphi_\sigma\varphi_\tau^*\bigr].
\end{equation}
Composition with a single-qubit Clifford gate on $L_1$ merely permutes the three non-identity Paulis and leaves the uniform measure on the Clifford group invariant. Consequently, the average $\mathbb E\varphi_\sigma=A_n$ is identical for every $\sigma$, and the pairwise averages $\mathbb E\varphi_\sigma\varphi_\tau^*=B_n$ ($\in \mathbb R$) are identical for every pair. The mean fidelity collapses to
\begin{equation}
\avg{\F}=\tfrac14+\tfrac38(B_n-\mathrm{Re}A_n).
\end{equation}
The image of $X$ is uniformly distributed over the $4^n -1$ non-identity Pauli strings, of which $\binom nk3^k$ have weight $k$; summing the corresponding phases supplies the closed form for $A_n$. The ordered pair of images ($P_X, P_Z$) is uniformly distributed over the set of pairs that anticommute on an odd number of sites. On a single site, the sixteen possible pairs contribute to the generating function $e^{ig(x-z)}$ (where $x,z \in \{0,1\}$ indicate non-identity entries); their sum is $s_0=10+6\cos g$. Weighting the six anticommuting pairs by an extra minus sign produces the second generating function $s_1=6\cos g-2$. The odd-parity sector is therefore $(s_0^n-s_1^n)/2$ while the total number of pairs is $(16^n-4^n)/2$, giving $B_n$.

\paragraph{Asymptotics.}
Substitute $g=\gamma/n$. The relevant logarithms expand as
\begin{align}
n\ln\Bigl[\frac{1+3e^{ig}}{4}\Bigr]&=\tfrac34 i\gamma-\tfrac{3\gamma^2}{32n}+O(n^{-2}),\\
n\ln\Bigl[\frac{10+6\cos g}{16}\Bigr]&=-\tfrac{3\gamma^2}{16n}+O(n^{-3}),
\end{align}
while all remaining terms decay exponentially with $n$. At the asymptotically optimal coupling $\gamma = 4 \pi/3$, one recovers the finite-size expansion
\begin{equation}
1-\avg{\F}=\tfrac{27\gamma^2}{256n}+O(n^{-2})=\tfrac{3\pi^2}{16n}+O(n^{-2}).
\end{equation}
For the magic of the coupling itself, a unitary of the form $e^{-i \theta K}$ acting on the $m$ qubits has Choi-state SRE $\tfrac{4\theta^2}{d_m^2\ln2}\bigl(d_m\mathrm{tr}K^2-(\mathrm{tr}K)^2\bigr)+O(\theta^4)$ with $d_m=2^m$. (The formula follows from the orthogonality of the Pauli transfer matrix together with the identity $\sum_P PKP=d_m\,\mathrm{tr}(K)\openone$.) Each local factor of the coupling Eq.~\eqref{eq:coupling} is, up to a global phase, equal to $e^{-ig\Pi_i}$ with $\mathrm{tr}\,\Pi_i=\mathrm{tr}\,\Pi_i^2=1$ and $d_m=4$, and therefore contributes $3g^2/(4\ln2)$. Evaluating all $256$ Pauli expectation values of the Choi state of a single factor gives its exact SRE, $-\log_2[(89+36\cos2g+3\cos4g)/128]$, whose expansion reproduces $3g^2/(4\ln2)$. The $n$ factors act on disjoint pairs of qubits, which gives Eq.~\eqref{eq:magicexact} and, at $g=4\pi/(3n)$, Eq.~\eqref{eq:magiccoupling}.

\paragraph{Fluctuations.}
For arguments of order $1/n$, $n\ln(t_+/16)=\tfrac34 in(\alpha+\beta+\delta)-\tfrac{3n}{32}(\alpha^2+\beta^2+\delta^2)+\dots$, while $(t_-/16)^n$ is exponentially small, so $(w_X,w_Y,w_Z)$ are asymptotically independent Gaussians with mean $3n/4$ and variance $3n/16$, and $\delta_\sigma\equiv gw_\sigma-3\gamma/4$ has variance $\sigma^2=3\gamma^2/16n$. At generic $\gamma$, $\F=\F_\infty+\tfrac18\sin(3\gamma/4)\sum_\sigma\delta_\sigma+O(\delta^2)$, which gives Eq.~\eqref{eq:varcliff}. At $\gamma=4\pi/3$ the linear term vanishes and $1-\F=\delta^{\mathsf T}\!A\,\delta+O(\delta^3)$ with $A=\tfrac14\openone-\tfrac1{16}J$, where $J$ is the all-ones $3\times3$ matrix. Gaussian moments then give $\mathbb E(1-\F)=\sigma^2\,\mathrm{tr}A=\tfrac9{16}\sigma^2$ and $\mathrm{Var}\,\F=2\sigma^4\,\mathrm{tr}A^2=\tfrac{33}{128}\sigma^4=11\pi^4/384n^2$, using $\sigma^2=\pi^2/3n$.

\paragraph{Parity at $g=\pi$}
On each site, the weight indicators obey the linear relation $y \equiv x + z +a \pmod 2$, where $a$ marks an anticommuting pair. Summing over all sites and using the fact that the total number of anticommuting sites is odd yields the global parity constraint $w_Y \equiv w_X+w_Z+1$. Consequently, the triple ($w_X,w_Y,w_Z$) is either entirely odd (in which case $\F=1$) or contains exactly one odd entry (in which case $\F=0$). At the special value $g=\pi$, one has the additional identity $B_n=-A_n$, and the closed-form average Eq.~\eqref{eq:closed} reduces to the fraction Eq.~\eqref{eq:pi}.

\section{Numerical methods}
\label{app:numerics}
Qubits are ordered as $\text{ref}, M, L_1,\ldots,L_n,R_1,\ldots,R_n$ and pure states are stored as dense complex tensors. Uniformly random Clifford unitaries are sampled with the stim library~\cite{gidney2021stim}; for $n=2$, the entire group of $11520$ elements is enumerated exhaustively. Haar-random unitaries are generated by QR-decomposition followed by the phase correction of Mezzadri~\cite{mezzadri2006generate}. The SRE is evaluated exactly via a fast Walsh-Hadamard transform~\cite{georges2025pauli}. Doped circuits insert a single $T$ gate on a uniformly chosen qubit between successive Clifford layers. The Haar variances of Fig.~\ref{fig:main}(b) use $2000$, $2000$, $1500$, $1000$, $600$, $300$, and $150$ scramblers at $n=2,\dots,8$, with errors from $300$ bootstrap resamples; the exact Clifford moments were checked against exhaustive enumeration of all anticommuting pairs for $2\le n\le5$. The SYK control of Sec.~\ref{sec:numerics} uses exact sparse state-vector evolution with $N_{\rm maj}=10$ Majorana modes per side at $\beta=2J^{-1}$, coupling strength $g=5.76$, and $28$ extraction times uniformly spaced in the interval $[0,18]$.

\end{widetext}
\bibliographystyle{apsrev4-2}
\bibliography{MBL}
\end{document}